\documentclass[review,number,sort&compress]{elsarticle}

\makeatletter
\def\ps@pprintTitle{%
 \let\@oddhead\@empty
 \let\@evenhead\@empty
 \def\@oddfoot{\centerline{\thepage}}%
 \let\@evenfoot\@oddfoot}
\makeatother

\usepackage{tikz}
\usepackage{float}
\usepackage{graphicx}
\usepackage{gensymb}
\usepackage{amssymb}
\usepackage{color,soul}
\usepackage{url}
\usepackage{ulem}
\usepackage{textcomp}
\usepackage[figuresright]{rotating}
\usepackage{subcaption}
\usepackage{lineno}
\usepackage{hyperref}
\usepackage{cleveref}
\usepackage{libertine}
\usepackage{siunitx}
\usepackage[printwatermark]{xwatermark}
\usepackage[a4paper, total={5.0in, 8in}]{geometry}
\usepackage[T1]{fontenc}

\definecolor{amethystbg}{rgb}{0.6, 0.4, 0.8}
\definecolor{coolgreybg}{rgb}{0.55, 0.57, 0.67}
\definecolor{babypinkbg}{rgb}{0.96, 0.76, 0.76}
\definecolor{cadmiumgreenbg}{rgb}{0.0, 0.42, 0.24}
\definecolor{bluebg}{rgb}{.63,.79,.95}
\definecolor{orangebg}{rgb}{1,0.5,0}
\colorlet{lightbluebg}{bluebg!40}
\colorlet{lightorangebg}{orangebg!40}
\colorlet{lightcadmiumgreenbg}{cadmiumgreenbg!40}
\colorlet{lightbabypinkbg}{babypinkbg!40}
\colorlet{lightcoolgreybg}{coolgreybg!40}
\colorlet{lightamethystbg}{amethystbg!40}

\DeclareGraphicsExtensions{.eps,.pdf,.png,.jpg,}

\begin{document}

\newcommand{\etal}{{\it et~al.}}
\newcommand{\geant} {{{G}\texttt{\scriptsize{EANT}}4}}
\newcommand{\srim} {\texttt{SRIM}}
\newcommand{\python} {\texttt{Python}}
\newcommand{\pandas} {\texttt{pandas}}
\newcommand{\SciPy} {\texttt{SciPy}}
\newcommand{\ROOT} {\texttt{ROOT}}

\DeclareRobustCommand{\hlb}[1]{{\sethlcolor{lightbluebg}\hl{#1}}}
\DeclareRobustCommand{\hlo}[1]{{\sethlcolor{lightorangebg}\hl{#1}}}
\DeclareRobustCommand{\hlg}[1]{{\sethlcolor{lightcadmiumgreenbg}\hl{#1}}}
\DeclareRobustCommand{\hlp}[1]{{\sethlcolor{lightbabypinkbg}\hl{#1}}}
\DeclareRobustCommand{\hlgr}[1]{{\sethlcolor{lightcoolgreybg}\hl{#1}}}
\DeclareRobustCommand{\hla}[1]{{\sethlcolor{lightamethystbg}\hl{#1}}}

\begin{frontmatter}


\title{\geant-based calibration of an organic liquid scintillator\tnoteref{label1}}

	\author[lund]{N.~Mauritzson}
	\author[lund]{K.G.~Fissum\corref{cor1}}
	\author[lund]{H.~Perrey}
	\author[glasgow]{J.R.M.~Annand}
	\author[lund]{R.J.W.~Frost}
	\author[ess,glasgow,milan]{R.~Hall-Wilton}
	\author[ess,glasgow]{R.~Al~Jebali}
	\author[ess]{K.~Kanaki}
	\author[lund,ess]{V.~Maulerova-Subert\fnref{fn2}}
	\author[lund]{F.~Messi\fnref{fn3}}
	\author[lund]{E.~Rofors}

	\address[lund]{Division of Nuclear Physics, Lund University, SE-221 00 Lund, Sweden}
	\address[ess]{Detector Group, European Spallation Source ERIC, SE-221 00 Lund, Sweden}
	\address[glasgow]{SUPA School of Physics and Astronomy, University of Glasgow, Glasgow G12 8QQ, Scotland, UK} 
	\address[milan]{Dipartimento di Fisica ``G. Occhialini'', Universit\`a degli Studi di Milano-Bicocca, Piazza della Scienza 3, 20126 Milano, Italy}
	
	\tnotetext[label1]{The data set doi:10.5281/zenodo.5524234 is available for download from \href{https://zenodo.org/record/5524234}{https://zenodo.org/record/5524234.}}
	\cortext[cor1]{Corresponding author. Telephone:  +46 46 222 9677; Fax:  +46 46 222 4709}
	\fntext[fn2]{present address: CERN, European Organization for Nuclear Research, 1211 Geneva, Switzerland and Hamburg University, D-20148 Hamburg, Germany}
	\fntext[fn3]{present address: DVel AB, Scheelevägen 32, SE-223 63 Lund, Sweden}
	
\begin{abstract}
A light-yield calibration of an NE~213A organic liquid scintillator detector has been performed using both monoenergetic and polyenergetic gamma-ray sources. Scintillation light was detected in a photomultiplier tube, and the corresponding pulses were subjected to waveform digitization on an event-by-event basis. The resulting Compton edges have been analyzed using a \geant~simulation of the detector which models both the interactions of the ionizing radiation as well as the transport of scintillation photons. The simulation is calibrated and also compared to well-established prescriptions used to determine the Compton edges, resulting ultimately in light-yield calibration functions. In the process, the simulation-based method produced information on the gain and intrinsic pulse-height resolution of the detector. It also facilitated a previously inaccessible understanding of the systematic uncertainties associated with the calibration of the scintillation-light yield. The simulation-based method was also compared to well-established numerical prescriptions for locating the Compton edges. Ultimately, the simulation predicted as much as 17\% lower light-yield calibrations than the prescriptions. These calibrations indicate that approximately 35\% of the scintillation light associated with a given gamma-ray reaches the photocathode.  It is remarkable how well two 50 year old prescriptions for calibrating scintillation-light yield in organic scintillators have stood the test of time.
\end{abstract}

\begin{keyword}
scintillation light-yield calibration, organic liquid scintillator, NE~213A, gamma-rays, Compton edge, \geant
\end{keyword}

\end{frontmatter}

\section{Introduction}
\label{section:Introduction}

Due to relatively high detection efficiency, strong inherent gamma-ray rejection properties, and fast scintillation pulses, organic liquid scintillators are typically employed to detect fast (MeV) neutrons in mixed neutron and gamma-ray fields.  The aromatic organic liquid scintillator NE~213~\cite{ne213} was originally introduced in the 1960s~\cite{BATCHELOR196170} and poses a non-negligible health risk. However, the excellent intrinsic neutron/gamma-ray pulse-shape discrimination characteristics and high fast-neutron detection efficiency continue to make NE 213 an excellent choice for fast-neutron applications. In this paper, the scintillation-light yield of the more recent NE~213A version of the liquid is calibrated using the Compton edges in measured energy distributions from a set of gamma-ray sources. This effort has been undertaken as the first step in a systematic program of parametrizing the scintillation-light yields of some recently developed organics and oils. The analysis of the data has been greatly facilitated by a \geant~simulation of the detector apparatus which models the interactions of gamma-rays and secondary electrons as well as the scintillation photon transport.

\section{Apparatus}
\label{section:Apparatus}

\subsection{Gamma-ray sources}
\label{subsection:GammaRaySources}
Above gamma-ray energy $E_{\gamma}$~$\sim$~100\,keV, the scintillation-light yield produced in organic liquids by atomic electrons freed by interactions with incident gamma-rays is very close to linear~\cite{KNOX1972519,GFKNOLL}. The low average $Z$ value typical for organics results in the gamma-ray/electron interactions being dominated by Compton scattering. Above $E_{\gamma}$~$=$~ 1.022\,MeV, pair production takes over and dominates by $\sim$5\,MeV. Measured Compton edges located at energy $E_{\rm CE}$ may be evaluated to calibrate the scintillation-light yield of a detector. Table~\ref{table:GammaRaySources} summarizes the radioactive sources used in this work.

\begin{table}[H]
    \centering
    \caption{Gamma-ray sources. ``Single'' refers to sources where a single gamma-ray was considered, while ``double'' refers to sources where two gamma-rays were considered.}
    \begin{tabular}{l l l l} \hline
        Source     & $E_{\gamma}$ [MeV] & $E_{\rm CE}$ [MeV$_{ee}$] &  Type \\
        \hline
        $^{22}$Na  &               0.51 &                      0.34 & double \\ 
        $^{137}$Cs &               0.66 &                      0.48 & single \\
        $^{60}$Co  &               1.17 &                      0.96 & double \\
        $^{22}$Na  &               1.28 &                      1.06 & double \\ 
        $^{60}$Co  &               1.33 &                      1.12 & double \\
        $^{232}$Th &               2.62 &                      2.38 & single \\
        AmBe       &               4.44 &                      4.20 & single \\ \hline
    \end{tabular}
    \label{table:GammaRaySources}
\end{table}

\subsection{NE~213A liquid-scintillator detector}
\label{subsection:NE213Detector}

\noindent
The volatile, corrosive, toxic, pungent, xylene-based scintillator NE~213 has long served as the baseline organic liquid against which all other organics are judged. In this work, the derivative pseudocumene-based scintillator NE~213A was employed~\cite{ANNAND1997}. Table~\ref{table:NE213A_properties} presents some of the well-known properties of NE~213A.

\begin{table}[H]
    \centering
    \caption{Properties of NE~213A.}
    \begin{tabular}{l l}\hline
         Solvent                        &       Pseudocumene (C$_9$H$_{12}$) \\
         Flash point                    &        $\sim$54\,$^{\circ}$C \\
         Density                        &         $\sim$0.9 g/cm$^3$ \\
         Light output                   &     $\sim$75\% of anthracene (pristine) \\
         Decay times                    &          $\sim$3, $\sim$32, $\sim$270\,ns \\
         Wavelength of maximum emission &                 $\sim$420 nm \\ \hline
    \end{tabular}
    \label{table:NE213A_properties}
\end{table}

\begin{figure}[H] 
    \centering
    \includegraphics[width=125mm]{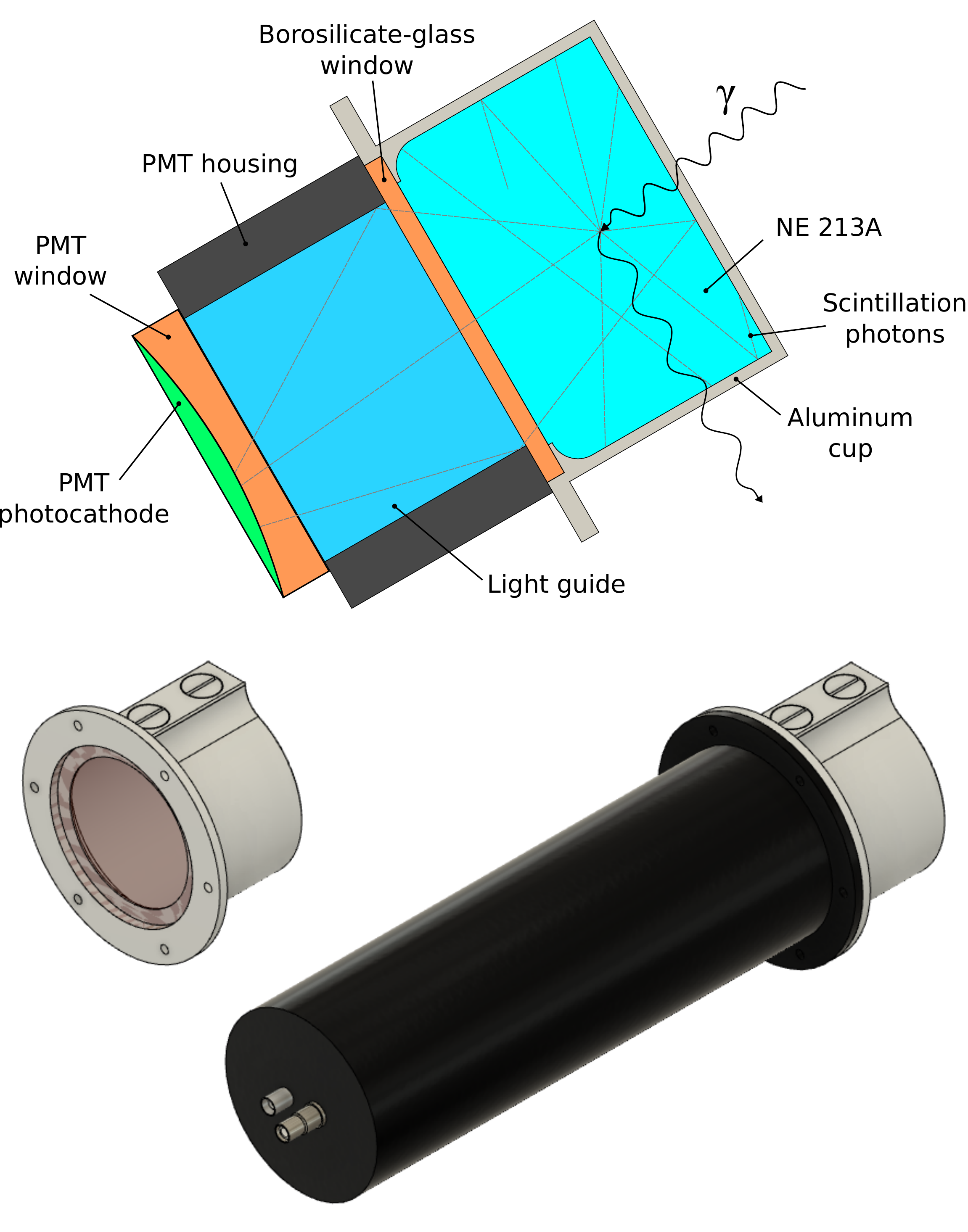}
    \caption{The NE~213A detector. Top: the cell, lightguide, PMT window, and photocathode. From the right, the NE~213A scintillator (cyan) was housed inside an aluminum cup (light gray) sealed with a borosilicate-glass window (orange). This window contacted a light guide (light blue) within the PMT housing (black), which in turn contacted the PMT window (orange) and photocathode (green). Middle left: oblique view of the scintillator cell. From the right, the cylindrical cup (light gray) and the circular borosilicate-glass window (light brown). The screws shown on top of the cup facilitated the filling. Bottom right: oblique view of the entire detector. From the right, the cell (light gray) and the $\mu$-metal shielded PMT and base housing (black). Contacts for signal and high voltage (gray) extend to the left from the base of the housing. For interpretation of the references to color in this figure caption, the reader is referred to the web version of this article.
    }
    \label{figure:ne213_detector}
\end{figure}

Figure~\ref{figure:ne213_detector} shows sketches of the liquid-scintillator detector. The scintillator housing was a 3~mm thick cylindrical aluminum cup 62\,mm deep by 94\,mm in diameter coated internally with the TiO$_2$-based reflective paint EJ~520~\cite{ej520}. A 5\,mm thick borosilicate glass optical window~\cite{borosilicate} was attached to the aluminum cell using Araldite 2000$+$ glue~\cite{araldite}. Together, the cup and the window formed a cell. A $\sim$430\,cm$^{3}$ volume of NE~213A was first flushed with nitrogen and then pushed into the cup using a pressurized nitrogen gas-transfer system. Viton O-rings~\cite{viton} were used to seal the filling penetrations. The filled cell was joined without any optical coupling medium to a 57\,mm long by 72.5\,mm diameter cylindrical lightguide made from PMMA UVT~\cite{pmma} coated externally with the TiO$_2$-based reflector EJ~510~\cite{ej510}. The cell/lightguide assembly was joined without any optical coupling medium to an ET type 9821K 3\,inch diameter photomultiplier tube (PMT) with a type B voltage divider~\cite{et_9821kb} equipped with a mu-metal magnetic shield and a spring to hold the PMT and PMMA faces in close contact.

\subsection{Signals, electronics, and data acquisition}
\label{subsection:ElectronicsAndDataAcquisition}

\begin{figure}[H] 
    \centering
    \includegraphics[width=1.\textwidth]{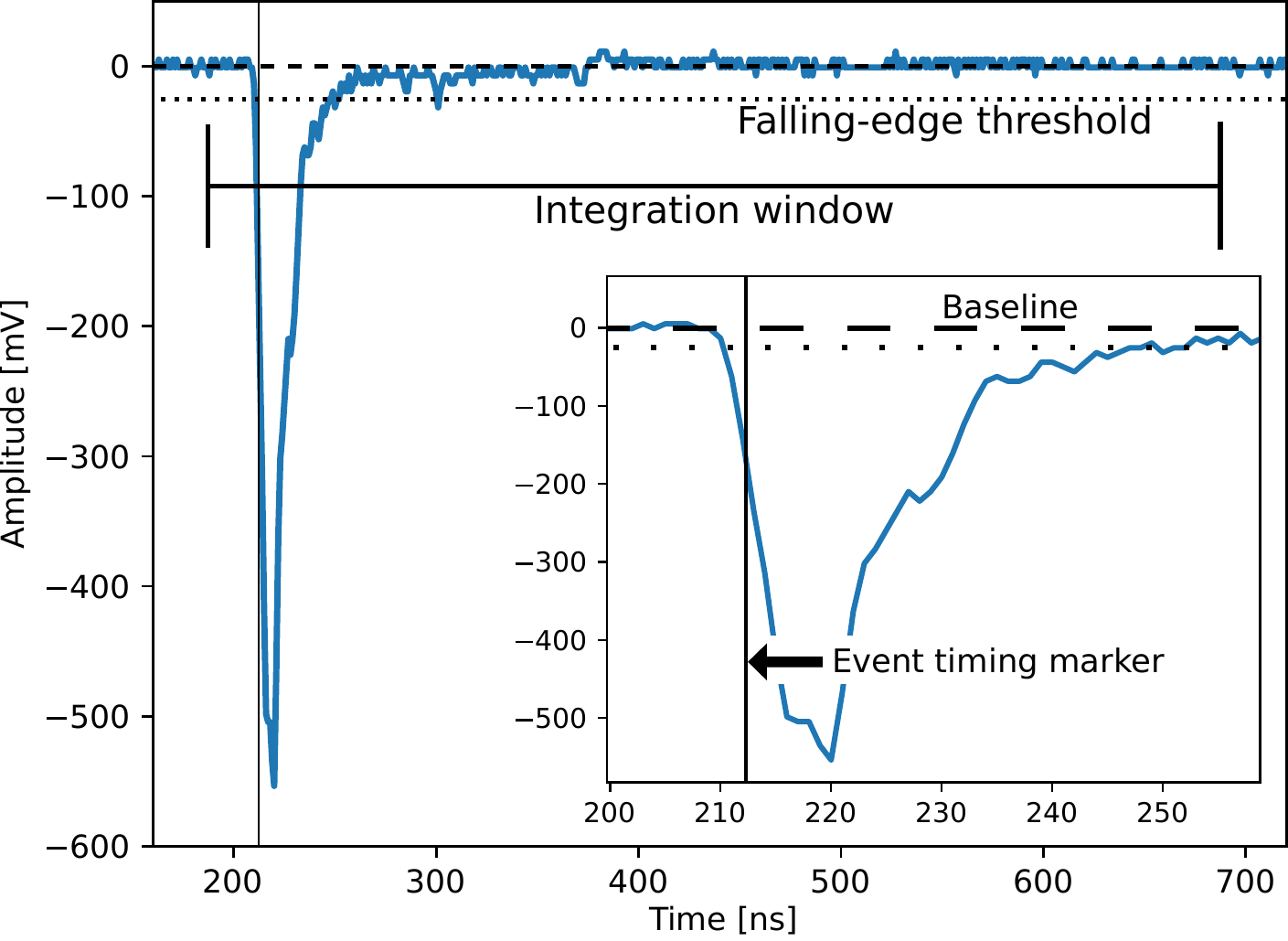}
    \caption{Signal waveform. The event pulse has a risetime of $\sim$5\,ns, an amplitude of $\sim$550\,mV and a falltime of $\sim$50\,ns. Illustrated are the event-timing marker, the 500\,ns integration window opening 25\,ns before the event-timing marker, and the $-$25\,mV falling-edge threshold. }
    \label{figure:waveform}
\end{figure}

The operating voltage of the detector was set at $-2$\,kV, a voltage employed for this detector in previous VME setups~\cite{SCHERZINGER201574, JEBALI2015102, JuliusScherzinger2016, SCHERZINGER201798, SCHERZINGER2017270}. At this voltage, a 1\,MeV$_{ee}$ signal had a risetime of $\sim$5\,ns, an amplitude of $\sim$900\,mV and a falltime of $\sim$60\,ns. The data-acquisition system was based on a CAEN VX1751 Waveform Digitizer~\cite{caen_vx1751} with a 10 bit ADC and an analog input bandwidth of 500\,MHz. The digitizer was configured for a 1\,\textmu{}s acquisition window with 10$^{9}$\,samples per second over a $-$1\,V dynamic input range. The voltage resolution was $\sim$1\,mV. In order to preserve the $-2$\,kV operating voltage used in the previous investigations, it was necessary to attenuate the analog signals from the detector by 16\,dB using a CAEN N858 dual attenuator module~\cite{caen_n858}. Figure~\ref{figure:waveform} shows a typical waveform. The internal falling-edge threshold was set to $-$25\,mV. The waveform of each pulse was analyzed using a suite of analysis software~\cite{nppp2020} developed in-house. Analysis of the data was performed using the \python-based~\cite{python} code libraries \pandas~\cite{pandas}, \SciPy~\cite{scipy}, and \texttt{numpy}~\cite{numpy}, where the signal baseline was first subtracted so that the charge corresponding to each scintillation pulse could be determined by integration. The event-timing marker was obtained using a standard zero-crossover method~\cite{GFKNOLL}. Voltage sampling was started 25\,ns before the event-timing marker and extended to 475\,ns after the event-timing marker. Integration was performed offline over this 500\,ns window which will be required for neutron/gamma-ray pulse-shape discrimination, resulting in an offline software-based charge-to-digital conversion. The conversion was calibrated to 6.35$\pm$5.5\% fC/QDC~channel using a charge-injection circuit.

\subsection{\geant~simulation}
\label{subsection:GeantSimulation}
The response of the detector to gamma-rays was simulated using a C++ Monte Carlo model developed with the \geant~toolkit~\cite{AGOSTINELLI2003250}. \geant~version 4.10.04~\cite{1610988} patch 03 (8 February 2019) was employed, with a physics list based on the hadronic class \texttt{FTFP\_BERT\_HP} and electromagnetic physics classes \texttt{G4EmStandardPhysics} and \texttt{G4EmExtraPhysics}, using a procedure similar to that reported in Ref.~\cite{BOYD2021165174}. The resulting model was used to simulate the gamma-ray response by modeling the gamma-ray interactions in the detector and tracking the secondary electrons and scintillation photons~\cite{g4prm} that they produced. The NE~213A scintillator was attributed a scintillation light-yield gradient of 1700 scintillation photons per MeV$_{ee}$ ($\sim$10\% of anthracene) and a Birks parameter of 0.126\,mm/MeV. This low scintillation light yield resulted from the work of Scherzinger~\etal~\cite{JuliusScherzinger2016} simulating a NE~213 filled detector. The reduced scintillation-light yield includes scintillator-aging effects, PMT gain, and PMT-aging effects. These photons were then tracked using the optical properties (refractive index, reflectivity, attenuation length) of the components, where they underwent scattering, absorption, and boundary transitions on the journey towards the photocathode. The optical surface model `glisur'~\cite{g4bad,gumplinger02} was used. The boundaries between the NE~213A scintillator and the borosilicate glass window of the cup, the glass window and the PMMA UVT lightguide, the lightguide and the PMT window, and the PMT window and the curved photocathode were all assumed to be polished resulting in specular reflection. Scintillation photons that penetrated into the photocathode were converted into photoelectrons (see below). A dielectric-to-dielectric interface was employed at each to account for refraction. In contrast, for the reflective-painted boundary between the scintillator and the aluminum cup as well as the external cylindrical surface of the lightguide, a dielectric-to-metal interface was employed. For the cup, the `metal' was attributed the optical properties of the reflective paint used. For the lightguide, a 110\,$\mu$m layer of the paint (corresponding to 3 coats, as per manufacturer specifications) was modeled. Surface irregularities in the paints were addressed using an optical surface model with a `SetPolish' parameter of 0.1. The dry-fitted boundaries between the cup window and the lightguide as well as the lightguide and the PMT window were taken to be air gaps of 100\,$\mu$m to account for surface non-planarities. Photon transmission was sensitive to the existence of the air gaps, but relatively insensitive to their widths. When the gap widths were varied from 100\,$\mu$m to 300\,$\mu$m, the scintillation-light yield varied by $\sim$1\%. At the photocathode, photoelectrons were generated based on the wavelength-dependent quantum efficiency~\cite{et_9821kb}, average $\sim$23\%. The PMT gain was defined as the scale factor necessary to match the simulated-photoelectron distributions to the measured spectra and was treated as a free parameter. Smearing was applied to match the simulated photoelectron distributions to the measured data in the vicinity of the Compton edge using a least-squares method (see below). It ranged from $\sim$23\% at 0.34 MeV$_{ee}$ ($^{22}$Na) to $\sim$12\% at 4.20 MeV$_{ee}$ (AmBe), with an inverse dependence on energy. It includes non-pointlike source, signal-propagation, and electronic noise effects and agrees well with that observed for a very similar detector by Scherzinger~\etal~\cite{JuliusScherzinger2016} 

\section{Measurement}
\label{section:Measurement}

The calibration sources were systematically placed in front of the NE~213A detector which was aligned so that the cylindrical symmetry axis of the detector pointed at the source. Sources with an activity below 1\,MBq ($^{22}$Na, $^{137}$Cs, $^{232}$Th) were placed at a distance of 45\,cm from the face of the unshielded detector, while the distance was increased to 200\,cm for sources with an activity above 1\,MBq ($^{60}$Co, AmBe). Hydrogen-rich materials were removed from the vicinity of the setup to minimize the production of 2.22\,MeV gamma-rays from neutron capture during the AmBe irradiations. A typical run time was 1 hour. Prior to data collection, background was investigated using a 1.5\,inch LaBr$^{3}$(Ce) gamma-ray detector. Gamma-rays from the de-excitations of $^{40}$K (1.46\,MeV) and $^{208}$Tl (2.61\,MeV, 583\,keV, 510\,keV) were observed. As count rates were on the order of a few 100\,Hz, deadtime was very low, so that the room background could be subtracted from the source measurement after a straightforward realtime normalization. 

\section{Results}
\label{section:Results}

\begin{figure}[H] 
    \centering
    \includegraphics[width=1.\textwidth]{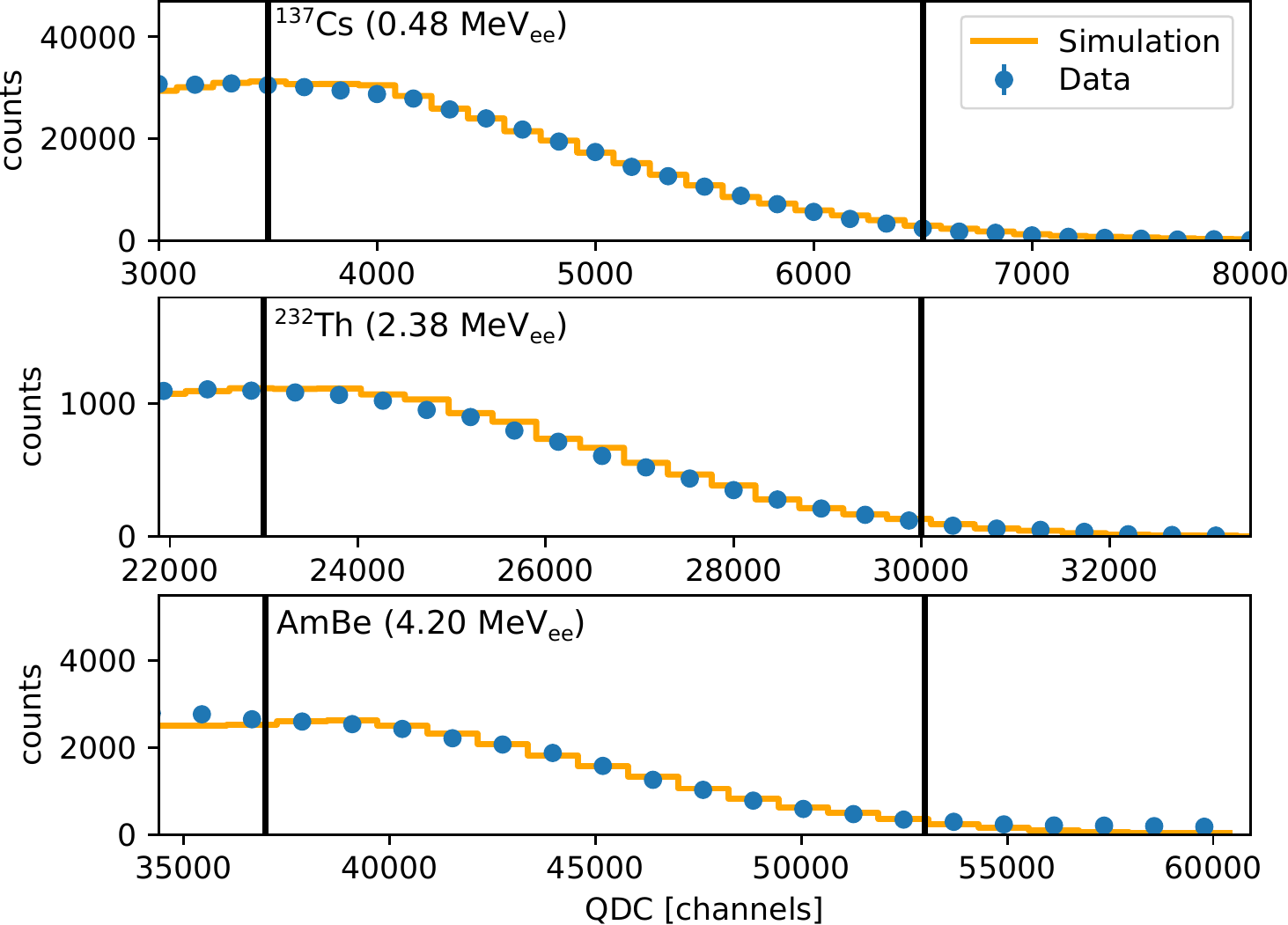}
    \caption{Replicating Compton edges. Data (filled points) and simulations (open histograms) are shown for three single-energy gamma-ray sources. The solid vertical bars in each panel denote the regions over which the agreement between the simulations and the data was optimized. The statistical uncertainties associated with the data points are smaller than the data points themselves.}
    \label{figure:methodsSpectra}
\end{figure}

\begin{figure}[H] 
    \centering
    \includegraphics[width=1.\textwidth]{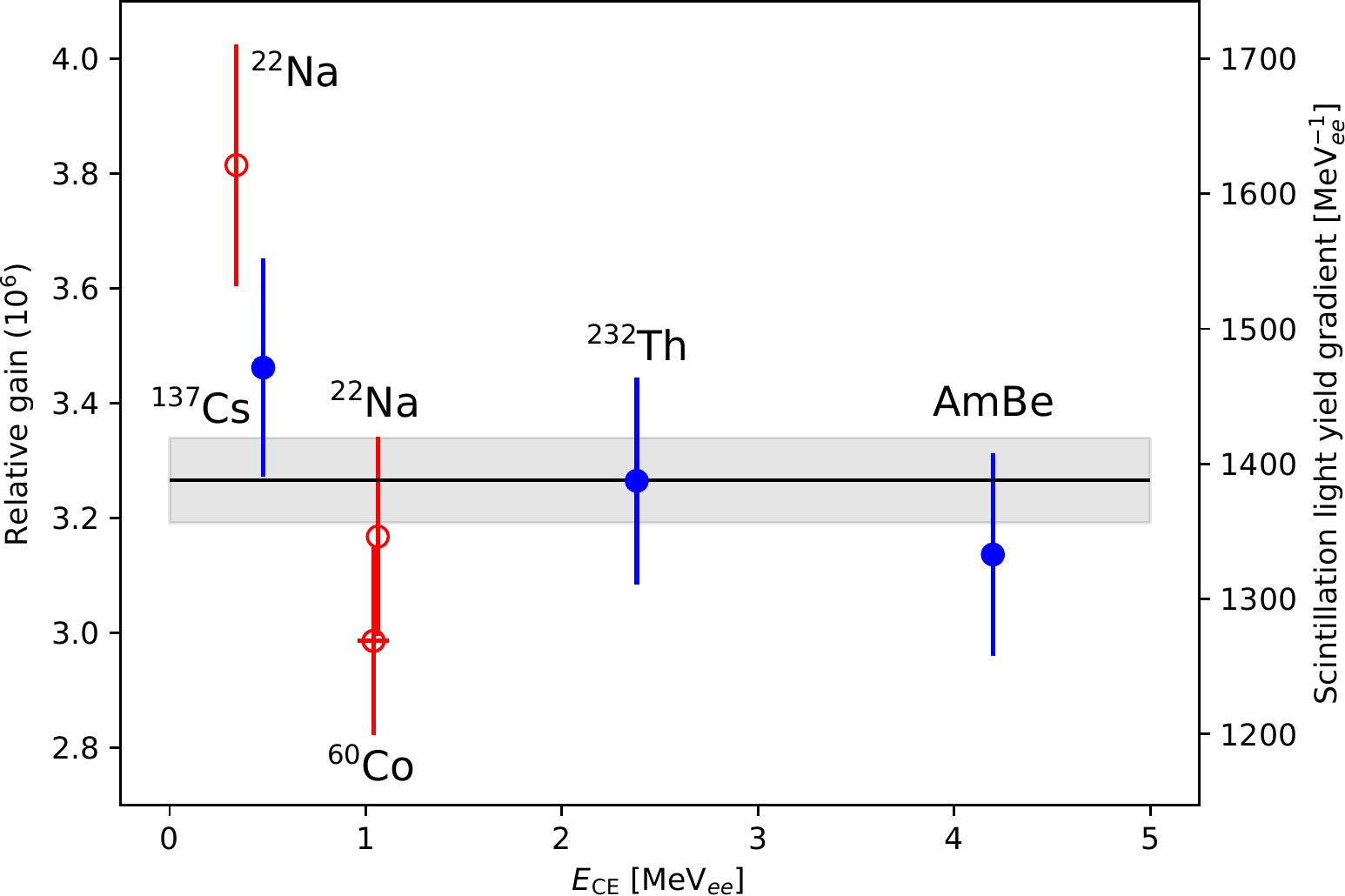}
    \caption{Inferred relative PMT gain. Gains extracted from matching the \geant~simulations to the Compton edges as detailed in Fig.~\ref{figure:methodsSpectra} using the 1700 scintillation photon per MeV$_{ee}$ light-yield gradient. Left axis, photoelectron multiplication, right axis, scintillation-light yield. Gains determined from both single-energy (filled circles) and double-energy (open circles) gamma-ray sources are shown (Table~\ref{table:GammaRaySources}). The error bars on the data points are dominated by the uncertainty in the QDC charge calibration.  The solid line indicates the average relative gain, while the uncertainty in the average value is represented by the shaded band.}
    \label{figure:Gains}
\end{figure}

Figure~\ref{figure:methodsSpectra} compares the \geant~simulations and the data in the vicinity of the Compton edges measured from three different sources, each emitting a single, well-defined gamma-ray.  As previously mentioned, the gain of the PMT was treated as a free parameter (see Fig.~\ref{figure:Gains}), and the simulated photoelectron distribution was matched to the measured data by applying an additional phenomenological smearing, all within a least-squares minimization. Agreement between the simulation and the data for each of the sources is excellent.

Figure~\ref{figure:Gains} shows the relative PMT gain inferred from matching the \geant~simulations obtained with the 1700 scintillation photon per MeV$_{ee}$ light-yield gradient and the offline QDC calibration to the data in the vicinity of the Compton edges. The gain of the PMT was taken to be $4~\cdot~10^6$ as per the data sheet. The two close-lying peaks from $^{60}$Co are shown as a single data point at an energy of 1.25\,MeV$_{ee}$ with an uncertainty (the horizontal error bar) of 80\,keV$_{ee}$. The uncertainty in the average relative gain has been taken from the uncertainty produced by the least-squares fitting algorithm. Over the $\sim$5\,MeV$_{ee}$ energy region investigated here, an average relative gain of (3.27~$\pm$~0.07)~$\cdot~10^6$, corresponding to (1388~$\pm$~31)~scintillation photons per MeV$_{ee}$, does a very good job of representing the results. This average gain corresponds to roughly 80\% of the 1700 scintillation photon per MeV$_{ee}$ used in the \geant~model. 

\begin{figure}[H] 
    \centering
    \includegraphics[width=1.\textwidth]{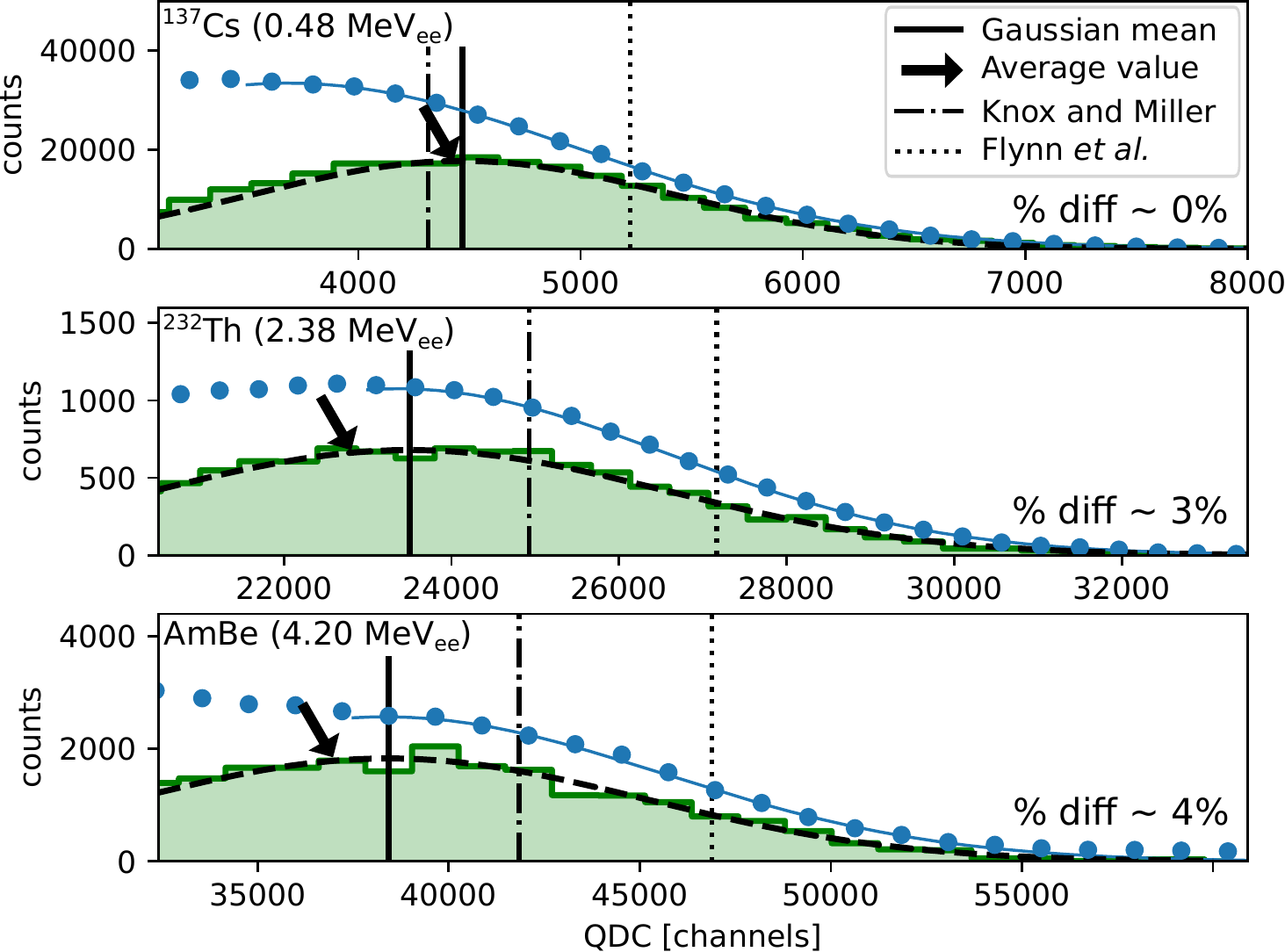}
    \caption{Scintillation-light yields for three single-energy gamma-ray emitters. Data (filled points) are shown for three single-energy gamma-ray sources together with simulations (shaded histograms) having a restrictive Compton-edge cut. Vertical solid lines illustrate the means of Gaussian distributions (dashed black curves) fitted to the simulated Compton-edge locations. The average values of the shaded histograms are indicated with angled arrows. Vertical dot-dashed lines (Knox and Miller, leftmost) and dotted lines (Flynn~\etal, rightmost) illustrate the Compton-edge locations extracted from the Gaussian functions fitted directly to the data over the optimization region (thin blue line). The statistical uncertainties associated with the data points are smaller than the data points themselves. For interpretation of the references to color in this figure caption, the reader is referred to the web version of this article.}
    \label{figure:methodsSpectra2}
\end{figure}

Figure~\ref{figure:methodsSpectra2} re-presents the data shown in Fig.~\ref{figure:methodsSpectra}, but this time in the context of Compton-edge analyses. Here, the well-known prescriptions for the locations of the Compton edges of Knox~and~Miller~\cite{KNOX1972519} and Flynn~\etal~\cite{FLYNN196413} are applied directly to the data. Both require that a Gaussian function is fitted to the high-energy side of the measured Compton edge. Flynn~\etal~associate the location of the half height of the distribution with 104\% of $E_{\rm CE}$ while Knox~and~Miller associate 89\% of the full height with 100\% of $E_{\rm CE}$. Note that with more recent input from Monte Carlo simulations, it has become generally accepted that these prescriptions are approximations to the actual location of the Compton edge~\cite{BEGHIAN196534,DIETZE1982549,ARNEODO1998285,MATEI2012135,JuliusScherzinger2016}. Also shown are the \geant~simulations. For each, the same individual PMT gains and phenomenological smearings used to produce Fig.~\ref{figure:methodsSpectra} have been employed. Further, a very restrictive cut where only those events with the recoiling electron receiving within 2\,keV of the maximum Compton-edge energy has been applied, resulting in an almost pure Compton-edge simulated dataset. Gaussian functions were fitted to the entire simulated distributions and the widths of these functions were used to determine the $\pm$3$\sigma$ event-summing region used for the calculation of the average peak position. The fitted Gaussians demonstrated the existence of tails in the distributions to lower QDC channels. These tails were energy dependent, ranging from $\sim$0\% of the integrated distribution below 1.12\,MeV$_{ee}$ to $\sim$14\% at 4.20\,MeV$_{ee}$. The tails resulted in an energy-dependent percentage difference between the fitted Gaussian mean and the average peak position of up to $\sim$4\% at 4.20\,MeV$_{ee}$. The locations of the simulated Compton-edge peaks relative to the locations of the Compton edges predicted by the Knox~and~Miller and Flynn~\etal~prescriptions are not constant offsets. They vary as a function of gamma-ray energy.

\begin{figure}[H] 
    \centering
    \includegraphics[width=1.\textwidth]{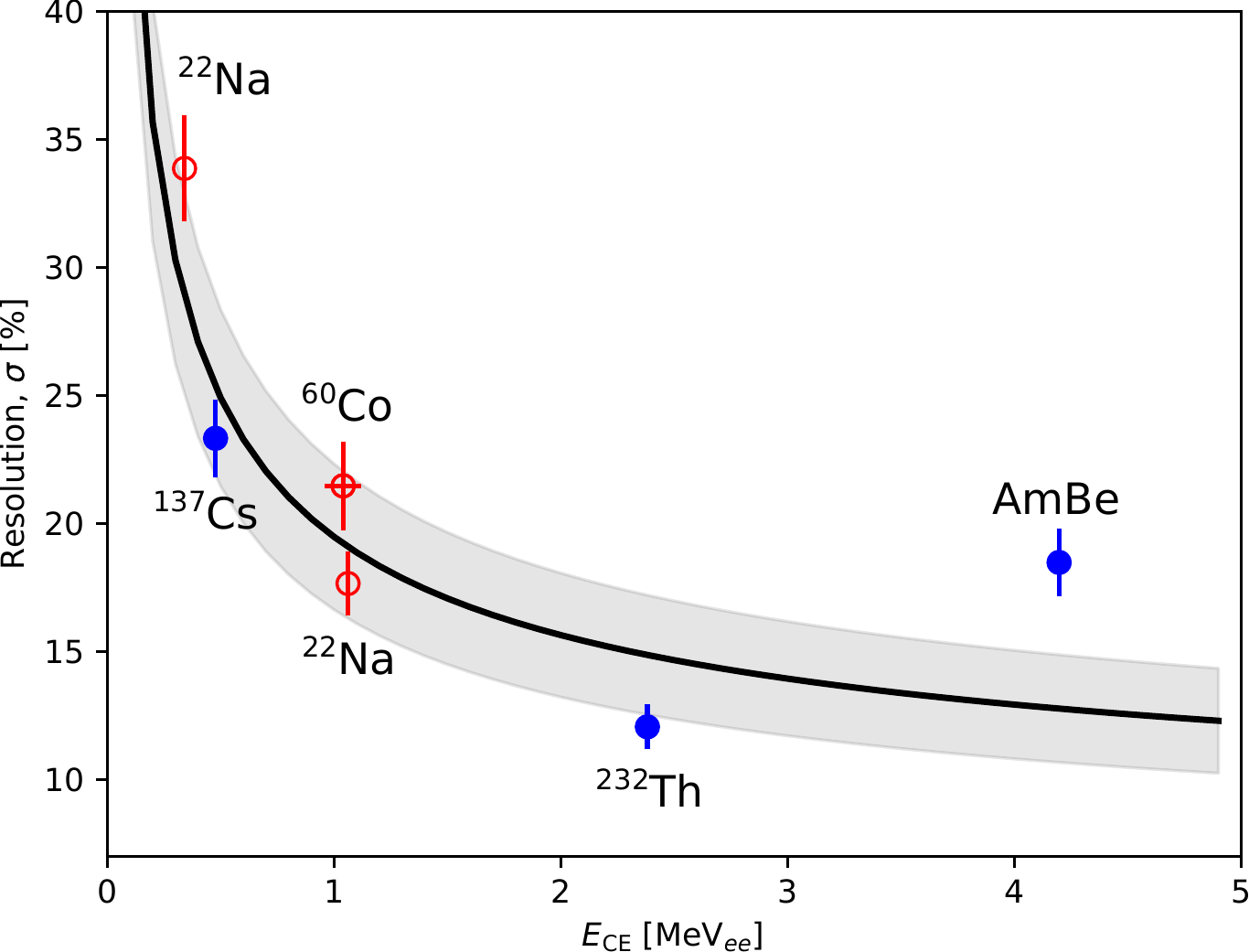}
    \caption{Resolution ($\sigma$). The energy resolution extracted from Gaussian functions fitted to the restrictively cut \geant~simulations. Results for single-energy (filled symbols) and double-energy (open symbols) gamma-ray sources are shown. The two close-lying peaks from $^{60}$Co are shown as a single data point as in Fig.~\ref{figure:Gains}. The error bars on the data points correspond to the quadratic sum of the uncertainties in the mean value and deviation of the fitted functions and the gains. A fitted $1/\sqrt{E_{\rm CE}}$ trend (solid line) is also shown. The uncertainty in this trend is represented by the shaded band.}
    \label{figure:EnergyResolution}
\end{figure}

Figure~\ref{figure:EnergyResolution} presents the intrinsic detector resolution extracted from the results of the \geant~simulations. The restrictive 2\,keV full Compton-edge energy cut is in place. If scintillation-photon statistics dominates the falloff in the energy resolution, the $1/\sqrt{E_{\rm CE}}$ dependence shown is anticipated.

\begin{figure}[H] 
    \centering
    \includegraphics[width=1.\textwidth]{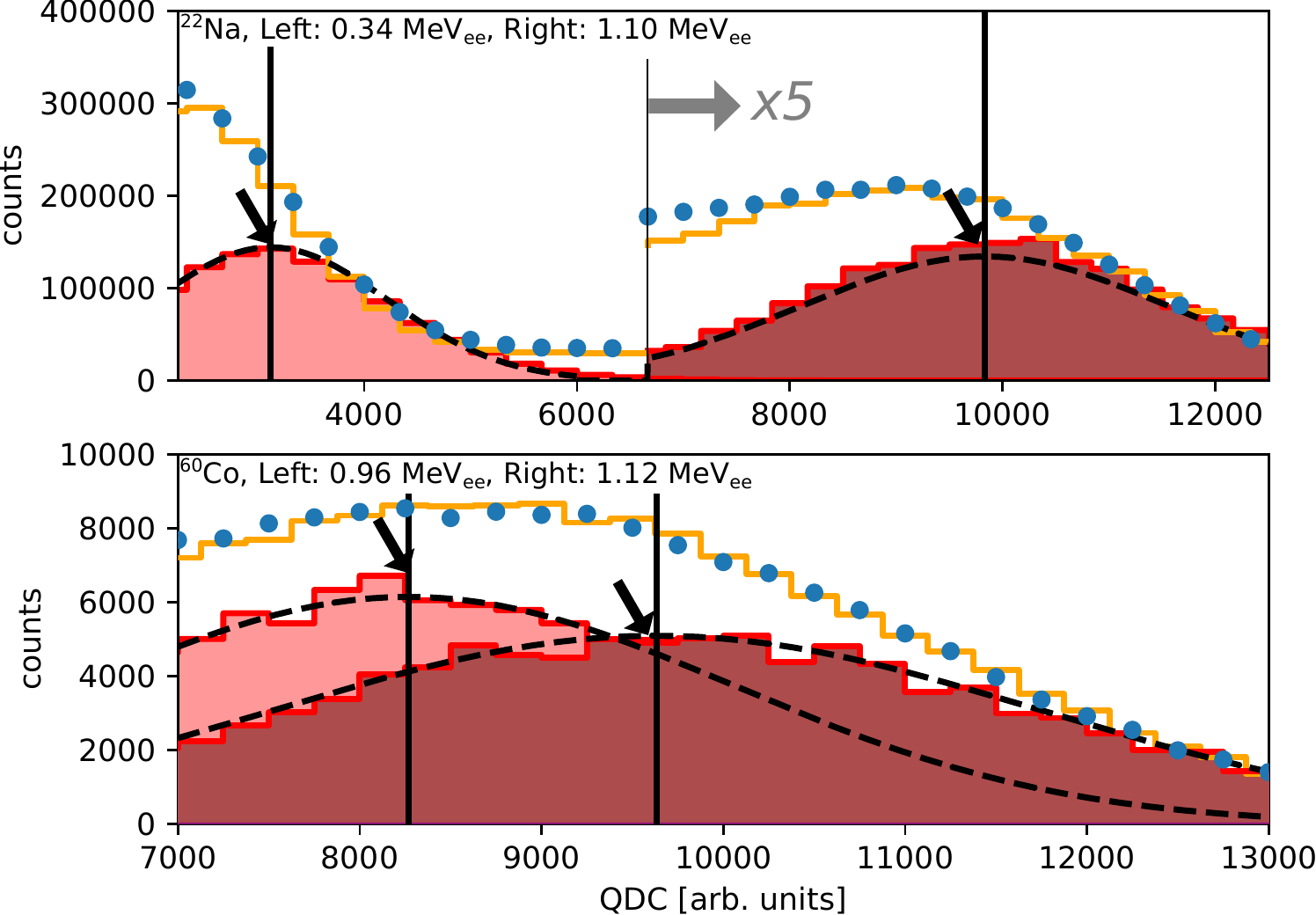}
    \caption{Scintillation-light yield for multiple gamma-ray emitters. Data (filled points), gain-matching simulations (open histograms) and very restrictively cut Compton-edge histograms (shaded) are shown. Vertical solid lines illustrate the means of Gaussian distributions (dashed black curves, drawn to guide the eye) fitted to the simulated Compton-edge locations. The average values are indicated with angled arrows. The statistical uncertainties associated with the data points are smaller than the data points themselves. Note the $\times$5 in the middle of the upper panel.}
    \label{figure:simulation}
\end{figure}

Figure~\ref{figure:simulation} shows the application of the simulation-based calibration to non-monoenergetic sources. Results obtained for $^{22}$Na and $^{60}$Co, two sources each emitting two relatively close-lying gamma-rays (759\,keV separation for $^{22}$Na and 160\,keV for $^{60}$Co) are shown. The $\sim$22\% energy resolution of the detector at these energies renders the methods of Knox and Miller and Flynn~\etal~difficult to apply, especially in the case of $^{60}$Co. By employing the \geant~based-calibration method, measured spectra obtained with these sources may be interpreted in a relatively straightforward manner. In each case, a single simulation of the source employing well-known gamma-ray branching ratios was performed in exactly the method described earlier. For $^{22}$Na, 90.2\% of decays yield both a 1.27\,MeV gamma-ray and a 511\,keV gamma-ray whereas 9.7\% yield only the 1.27\,keV gamma-ray. For $^{60}$Co, 99.88\% of decays yield both a 1.17\,MeV and a 1.33\,MeV gamma-ray whereas 0.12\% of decays yield only a 1.33\,MeV gamma-ray. Agreement between the branched double-energy gamma-ray simulations and the data is again excellent. The corresponding gains and resolutions have already been reported as open circles in Figs.~\ref{figure:Gains} and ~\ref{figure:EnergyResolution}, respectively.

\begin{figure}[H] 
    \centering
    \includegraphics[width=1.\textwidth]{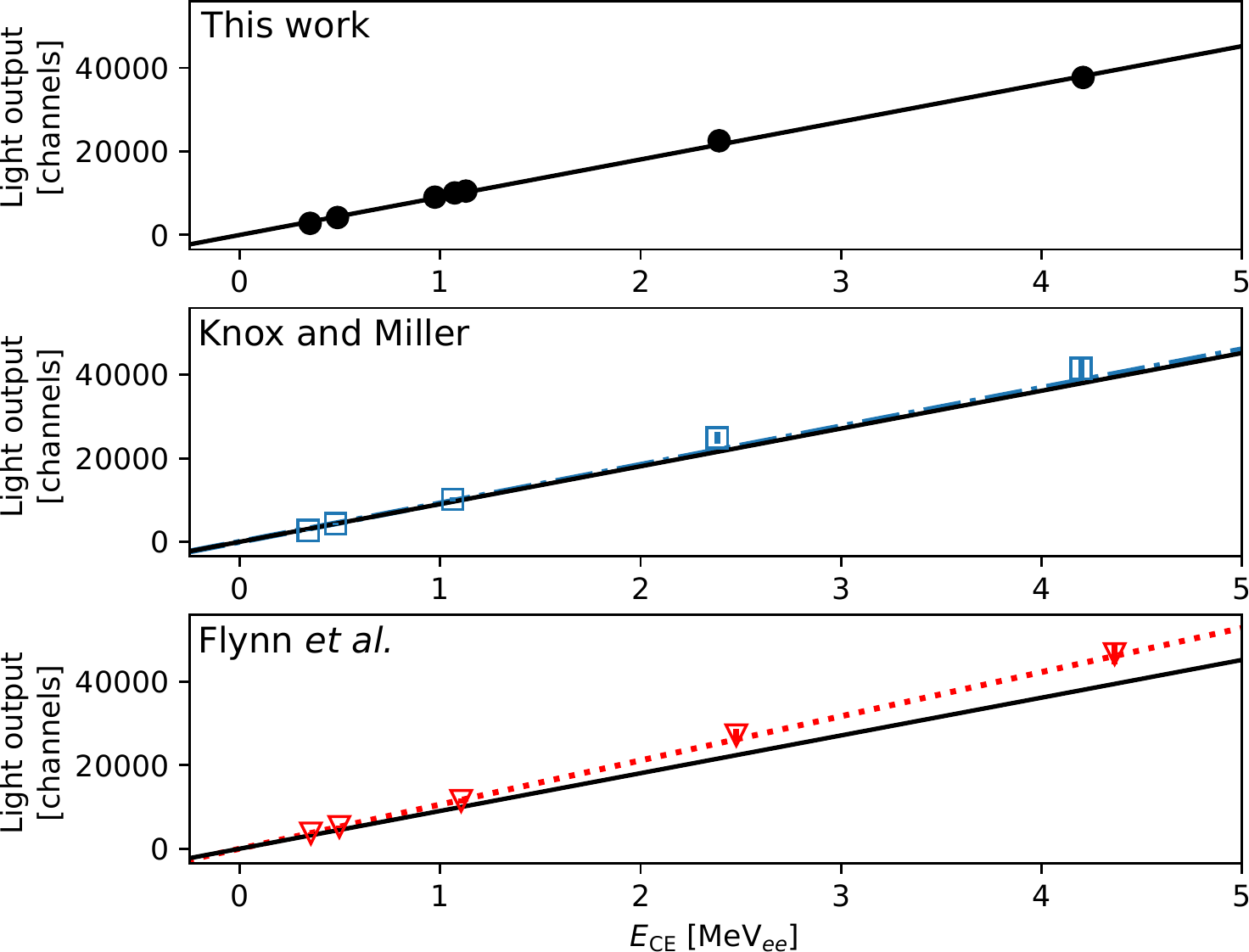}
    \caption{Light-output calibrations. Top panel: This work, average values (filled circles, solid line) of the restrictively cut \geant~simulations. The solid average-value line shown in this panel appears in all panels to facilitate comparison between approaches. Middle panel: Knox and Miller approach (open squares, dot-dashed line). Bottom panel: Flynn~\etal~approach~(open triangles, dotted line). Due to insufficient detector energy resolution, $^{60}$Co results are not shown for the Knox and Miller or Flynn~\etal~analyses. The uncertainties are smaller than the data points themselves.}
    \label{figure:FinalCalibrationPlot}
\end{figure}

Figure~\ref{figure:FinalCalibrationPlot} presents light-output calibrations obtained with linear fits to the data which are summarized in Table~\ref{tab:QDC_calibrations}. The fitted functions shown have been constrained to pass through the origin. The fits do a very good job of linearly replicating the simulated Compton-edge locations as a function of energy. The dominant systematic uncertainty in the data contributing to the uncertainty in the light-output calibration was the $\sim$5.5\% uncertainty in the charge calibration of the QDC. Systematic uncertainties arising from the analysis of the simulations included uncertainties arising from the fitted parameters ($<$3\%) the effects of the various cuts employed in the analysis. The uncertainty due to the various cuts ($<$1\%) was addressed by systematically varying the windows employed. No clear concensus regarding the systematic uncertainty associated with the Knox and Miller and Flynn~\etal~approaches exists. Systematic uncertainties arising from the fitted parameters and cuts, again $<$3\% and $<$1\% respectively, were consistent with those obtained with the \geant~simulations.

\begin{table}[H]
    \centering
    \caption{Linear fitted light-output calibrations.}
    \begin{tabular}{c|rrr|rr}
    \hline
                           & \multicolumn{3}{c|}{zero unenforced} & \multicolumn{2}{c}{zero enforced} \\
    \cline{2-6}
                    Method &                                    slope &    offset & $\chi_{\nu}^2$ &                                    slope & $\chi_{\nu}^2$ \\
                           & [$\frac{\rm channels}{\rm MeV_{\it ee}}$] & [channels] &          & [$\frac{\rm channels}{\rm MeV_{\it ee}}$] &          \\
    \hline
    \geant~(average value) &  9612 $\pm$ 326 &    $-$434 $\pm$ 203 & 0.4 &  9049 $\pm$ 190 & 1.3 \\
    \hline
    Knox and Miller        & 10443 $\pm$ 399 & $-$803 $\pm$ 215 & 0.4 &  9222 $\pm$ 229 & 5.1 \\
    Flynn~\etal~           & 10767 $\pm$ 426 & $-$147 $\pm$ 260 & 0.1 & 10576 $\pm$ 258 & 0.3 \\
    
    \hline
    \end{tabular}
    \label{tab:QDC_calibrations}
\end{table}

Table~\ref{tab:ScintillationLightYield} shows the number of scintillation photons per MeV$_{ee}$ reaching the photocathode unfolded for the three methods for linear fits where the line was constrained to pass through the origin. The Knox and Miller prescription is $\sim$2\% larger than the simulated average-value light yield. The Flynn~\etal~light prescription is $\sim$17\% larger than the simulated average-value light yield. Based upon the 1700 scintillation photons per MeV$_{ee}$ light-yield gradient employed in the \geant~simulation and the $\sim$80\% relative gain, $\sim$35\% of the scintillation light produced by a gamma-ray reaches the photocathode.

\begin{table}[H]
    \centering
    \caption{Scintillation photons reaching the photocathode.}
    \begin{tabular}{cc}
    \hline
                    Method & photons per MeV$_{ee}$ reaching the photocathode \\
    \hline
    \geant~(average value) & 483 $\pm$ 2.1\% \\
    \hline
    Knox and Miller        & 493 $\pm$ 2.5\% \\
    Flynn~\etal~           & 565 $\pm$ 2.4\% \\
    \hline
    \end{tabular}
    \label{tab:ScintillationLightYield}
\end{table}
 
\section{Summary and Discussion}
\label{Section:SummaryAndDiscussion}

A scintillation light-yield calibration of an NE~213A organic liquid scintillator detector (Fig.~\ref{figure:ne213_detector}) has been performed using single-energy and double-energy gamma-ray sources. An event-by-event waveform-digitization (Fig.~\ref{figure:waveform}) of the scintillation signals resulted in measured Compton-edge distributions. Interpretation of the Compton-edge distributions used a \geant-based simulation which models the interactions of ionizing radiation and the transport of scintillation photons produced along particle tracks. Simulations employed a 1700 photon per MeV$_{ee}$ light-yield gradient, tuned to the data using relative PMT gain as a scaling parameter, and matched to the Compton edges by applying an additional smearing (Fig.~\ref{figure:methodsSpectra}). The relative gain function determined in this manner was linear over the $\sim$5\,MeV$_{ee}$ range considered at a photoelectron multiplication of (3.27~$\pm$~0.07)~$\cdot~10^6$, corresponding to a (1388~$\pm$~31) scintillation photon per MeV$_{ee}$ light-yield gradient (Fig.~\ref{figure:Gains}). Charge distributions were determined as a function of electron energy by enforcing very strict cuts in the simulation around the upper edge of the recoiling electron energy-loss spectrum as well as considering the well-established prescriptions of Knox~and~Miller and Flynn~\etal~(Fig.~\ref{figure:methodsSpectra2}) These restricted simulated distributions facilitated an evaluation of the intrinsic detector resolution, which was determined to be $\sim$18\% at $\sim$1\,MeV$_{ee}$ and to fall off $\sim$1/$\sqrt{E_{\rm CE}}$ (Fig.~\ref{figure:EnergyResolution}). An advantage of the simulation approach over the prescriptions is that it allows for the unfolding of spectra from radioactive sources emitting more than one gamma-ray, even if the energy separation of the gamma-rays is small. To demonstrate this advantage, the entire simulation and analysis procedure was successfully repeated for two such sources, $^{22}$Na (759\,keV gamma-ray separation) and $^{60}$Co (160\,keV gamma-ray separation) (Fig.~\ref{figure:simulation}). Linear light-output calibrations were then determined (Fig.~\ref{figure:FinalCalibrationPlot}). The \geant-based method developed here was chosen as a benchmark. The prescriptions, while also linear, were $\sim$2\% (Knox and Miller) and $\sim$17\% larger (Flynn~\etal) than the benchmark. The functions indicate that $\sim$35\% of the scintillation light associated with a given gamma-ray reaches the photocathode. It is remarkable how well two 50 year old prescriptions for calibrating scintillation-light yield have stood the test of time.

\section*{Acknowledgements}
\label{Section:Acknowledgements}
Support for this project was provided by the European Union via the Horizon 2020 BrightnESS Project (Proposal ID 676548) and the UK Science and Technology Facilities Council (Grant No. ST/P004458/1).
\newpage

\bibliographystyle{elsarticle-num}


\begin{thebibliography}{00}

\bibitem{ne213} 
NE213 is no longer produced. Eljen Technologies EJ-301 (\url{http://www.eljentechnology.com/index.php/products/liquid-scintillators/71-ej-301} [accessed 2021, Sep. 24]) or Saint Gobain BC-501 (\url{https://www.crystals.saint-gobain.com/products/bc-501a-bc-519} [accessed 2021, Sep. 24]) are very similar.

\bibitem{BATCHELOR196170} 
R.~Batchelor et al.,
Nucl. Instr. and Meth. 13 (1961) 70.
\href{https://doi.org/10.1016/0029-554X(61)90171-9}{doi:10.1016/10.1016/0029-554X(61)90171-9.}

\bibitem{KNOX1972519} 
H.H.~Knox et al.,
Nucl. Instr. and Meth. 101 (1972) 519.
\href{https://doi.org/10.1016/0029-554X(72)90040-7}{doi:10.1016/0029-554X(72)90040-7.}

\bibitem{GFKNOLL} 
Radiation detection and measurement,
G.F.~Knoll,
2nd edition, Wiley, New York, U.S.A. (1989) 222,
ISBN: 9780471815044.

\bibitem{ANNAND1997} 
J.R.M.~Annand et al.,
Nucl. Instr. and Meth. in Phys. Res. A. 400, (1997) 344.
\href{https://doi.org/10.1016/S0168-9002(97)01021-8}{doi:10.1016/S0168-9002(97)01021-8.}

\bibitem{ej520} 
\url{http://www.ggg-tech.co.jp/maker/eljen/ej-520.html} [accessed 2021, Sep. 24].

\bibitem{borosilicate} 
See \url{http://www.us.schott.com/borofloat/english/index.html} [accessed 2021, Sep. 24]. Supplied
by Glasteknik i Emmaboda AB, Utv\"{a}gen 6 SE-361 31 Emmaboda, Sweden.

\bibitem{araldite} 
Araldite is a registered trademark of Huntsman. See \url{http://www.araldite2000plus.com} [accessed 2021, Sep. 24].

\bibitem{viton} 
Viton is a registered trademark of DuPont Performance Elastomers LLC.

\bibitem{pmma} 
Poly-methyl-methacrylate, also known as PMM, acrylic, plexiglass, and lucite. Supplied by Nordic Plastics Group AB, Bronsyxegatan 6, SE-213 75 Malmoe, Sweden.

\bibitem{ej510} 
See \url{https://eljentechnology.com/products/accessories/ej-510} [accessed 2021, Sep. 24].

\bibitem{et_9821kb} 
See \url{https://et-enterprises.com/images/data_sheets/9821B.pdf} [accessed 2021, Sep. 24].

\bibitem{SCHERZINGER201574} 
J.~Scherzinger et al.,
Appl. Radiat. Isop. 98 (2015) 74.
\href{https://doi.org/10.1016/j.apradiso.2015.01.003}{doi:10.1016/j.apradiso.2015.01.003.}

\bibitem{JEBALI2015102} 
R.~Jebali et al.,
Nucl. Instr. and Meth. in Phys. Res. A. 794, (2015) 102.
\href{https://doi.org/10.1016/j.nima.2015.04.058}{doi:10.1016/j.nima.2015.04.058.}

\bibitem{JuliusScherzinger2016} 
J.~Scherzinger et al.,
Nucl. Instr. and Meth. in Phys. Res. A. 840, (2016) 121.
\href{https://doi.org/10.1016/j.nima.2016.10.011}{doi:10.1016/j.nima.2016.10.011.}

\bibitem{SCHERZINGER201798} 
J.~Scherzinger et al.,
Appl. Radiat. Isop. 127 (2017) 98.
\href{https://doi.org/10.1016/j.apradiso.2017.05.014}{doi:10.1016/j.apradiso.2017.05.014.}

\bibitem{SCHERZINGER2017270} 
J.~Scherzinger et al.,
Appl. Radiat. Isop. 128 (2017) 270.
\href{https://doi.org/10.1016/j.apradiso.2017.05.022}{doi:10.1016/j.apradiso.2017.05.022.}

\bibitem{caen_vx1751} 
See \url{https://www.caen.it/products/vx1751/} [accessed 2021, Sep. 24].

\bibitem{caen_n858} 
See \url{https://www.caen.it/products/n858/} [accessed 2021, Sep. 24].

\bibitem{nppp2020} 
H.~Perrey et al., Nuclear Physics Pulse Processing Library,
available at \url{https://gitlab.com/ANPLU/nppp} [accessed 2021, Sep. 24].

\bibitem{python} 
G. van Rossum and F.L. Drake (editors), Python Reference Manual, PythonLabs (2001).
Python 3.8.5 available at \url{https://www.python.org/} [accessed 2021, Sep. 24].

\bibitem{pandas} 
The Pandas Development Team,
\pandas 1.2.3 available at \url{https://pandas.pydata.org/} [accessed 2021, Sep. 24],
\href{https://doi.org/10.5281/zenodo.4572994}{doi:10.5281/zenodo.4572994.}

\bibitem{scipy} 
\url{https://www.scipy.org/} [accessed 2021, Sep. 24],
\href{https://doi.org/10.5281/zenodo.4547611}{doi:10.5281/zenodo.4547611.}

\bibitem{mckinney10} 
W. McKinney, Proceedings of the 9th Python in Science Conference, 445 (2010) 56.
\href{https://doi.org/10.25080/Majora-92bf1922-00a}{doi:10.25080/Majora-92bf1922-00a.}

\bibitem{numpy} 
\texttt{numpy} 1.20.3 available at \url{https://pypi.org/project/numpy/} [accessed 2021, Sep. 24].

\bibitem{AGOSTINELLI2003250} 
S.~Agostinelli et al.,
Nucl. Instr. and Meth. in Phys. Res. A. 506, (2003) 250.
\href{https://doi.org/10.1016/S0168-9002(03)01368-8}{doi:10.1016/S0168-9002(03)01368-8.}

\bibitem{1610988} 
J.~Allison et al.,
IEEE Trans. Nucl. Sci. 53, (2006) 270.
\href{https://doi.org/10.1109/TNS.2006.869826}{doi:10.1109/TNS.2006.869826.}

\bibitem{g4prm} 
Geant4 physics reference manual (2020) Release 10.6 available at
\url{http://geant4-userdoc.web.cern.ch/geant4-userdoc/UsersGuides/PhysicsReferenceManual/fo/PhysicsReferenceManual.pdf}

\bibitem{g4bad} 
Geant4 book for application developers (2020) Release 10.6 available at
\url{http://geant4-userdoc.web.cern.ch/geant4-userdoc/UsersGuides/ForToolkitDeveloper/fo/BookForToolkitDevelopers.pdf}

\bibitem{gumplinger02} 
P. Gumplinger, Optical photon processes in Geant4, Users’ Workshop at CERN, Nov. 2002.

\bibitem{FLYNN196413} 
K.F.~Flynn et al.,
Nucl. Instr. and Meth. 27 (1964) 13.
\href{https://doi.org/10.1016/0029-554X(64)90129-6}{doi:10.1016/0029-554X(64)90129-6.}

\bibitem{BOYD2021165174} 
L.~Boyd et al.,
Nucl. Instr. and Meth. in Phys. Res. A. 998, (2021) 165174.
\href{https://doi.org/10.1016/j.nima.2021.165174}{doi:10.1016/j.nima.2021.165174.}

\bibitem{BEGHIAN196534} 
L.E.~Beghian et al.,
Nucl. Instr. and Meth. 35 (1965) 34.
\href{https://doi.org/10.1016/0029-554X(65)90004-2}{doi:10.1016/0029-554X(65)90004-2.}

\bibitem{DIETZE1982549} 
G.~Dietze et al.,
Nucl. Instr. and Meth. 139 (1982) 549.
\href{https://doi.org/10.1016/0029-554X(82)90249-X}{doi:10.1016/0029-554X(82)90249-X.}

\bibitem{ARNEODO1998285} 
F.~Arnedodo et al.,
Nucl. Instr. and Meth. in Phys. Res. A. 418, (1998) 285.
\href{https://doi.org/10.1016/S0168-9002(98)00679-2}{doi:10.1016/S0168-9002(98)00679-2.}

\bibitem{MATEI2012135} 
C.~Matei et al.,
Nucl. Instr. and Meth. in Phys. Res. A. 676, (2012) 135.
\href{https://doi.org/10.1016/j.nima.2011.11.076}{doi:10.1016/j.nima.2011.11.076.}

\end{thebibliography}

\end{document}